\documentclass[reprint,aps,prb,twocolumn,groupedaddress,nobibnotes,superscriptaddress]{revtex4-1}
\setcitestyle{numbers,square}

\usepackage[english]{babel}
\usepackage{amsmath,amssymb,amsfonts} 
\usepackage{graphicx}
\usepackage{bm}
\usepackage{xcolor}
\usepackage{color} 
\usepackage{dcolumn}
\usepackage{chngcntr}
\usepackage{hyperref}
\usepackage{mathtools}
\usepackage{amsmath}
\usepackage{amsfonts}
\usepackage{amsthm}
\usepackage{amssymb}
\usepackage{bm}
\usepackage{graphicx}
\usepackage{float}
\usepackage{multirow}
\usepackage{float}
\usepackage{placeins}
\usepackage{caption}
\usepackage{ragged2e}
\usepackage[mathlines]{lineno}
\modulolinenumbers[5]
\usepackage{commath}
\usepackage{verbatim}
\usepackage{filecontents}

\DeclareMathAlphabet{\mathsfit}{T1}{\sfdefault}{\mddefault}{\sldefault}
\SetMathAlphabet{\mathsfit}{bold}{T1}{\sfdefault}{\bfdefault}{\sldefault}
\usepackage[caption=false]{subfig}
\usepackage{makecell}

\begin{document}

\title{Vortices and backflow in hydrodynamic heat transport}

\author{Enrico Di Lucente}
\affiliation{Theory and Simulation of Materials (THEOS),
École Polytechnique Fédérale de Lausanne, Lausanne 1015, Switzerland}
\author{Francesco Libbi}
\affiliation{John A. Paulson School of Engineering and Applied Sciences, Harvard University, Cambridge, MA 02138, USA}
\author{Nicola Marzari}
\affiliation{Theory and Simulation of Materials (THEOS) and National Centre for Computational Design and Discovery of Novel Materials (MARVEL), École Polytechnique Fédérale de Lausanne, Lausanne 1015, Switzerland}
\affiliation{Laboratory for Materials Simulations, Paul Scherrer Institut, 5232 Villigen PSI, Switzerland}

\date{\today}

\begin{abstract}
Recent experiments have provided compelling and renewed interest in phonon hydrodynamics. At variance with ordinary diffusive heat transport, this regime is primarily governed by momentum-conserving phonon collisions. At the mesoscopic scale it can be described by the viscous heat equations (VHE), that resemble the Navier-Stokes equations (NSE) in the laminar regime. Here, we show that the VHE can be separated and recast as modified biharmonic equations in the velocity potential and stream function—solvable analytically. These two can be merged into a complex potential defining the flow streamlines, and give rise to two distinct temperature contributions, ultimately related to thermal compressibility and vorticity. The irrotational and incompressible limits of the phonon VHE are analyzed, showing how the latter mirrors the NSE for the electron fluid. This work also extends to the electron compressible regime that arises when drift velocities can be higher than plasmonic velocities. Finally, by examining thermal flow within a 2D graphite strip device, we explore the boundary conditions and transport coefficients needed to observe thermal vortices and negative thermal resistance, or heat backflow from cooler to warmer regions. This work provides novel analytical tools to design hydrodynamic phonon flow, highlights its generalization for electron hydrodynamics, and promotes additional avenues to explore experimentally such fascinating phenomena. 
\end{abstract}

\maketitle

\begin{figure}[!htb]
\centering
\includegraphics[width=0.48\textwidth]{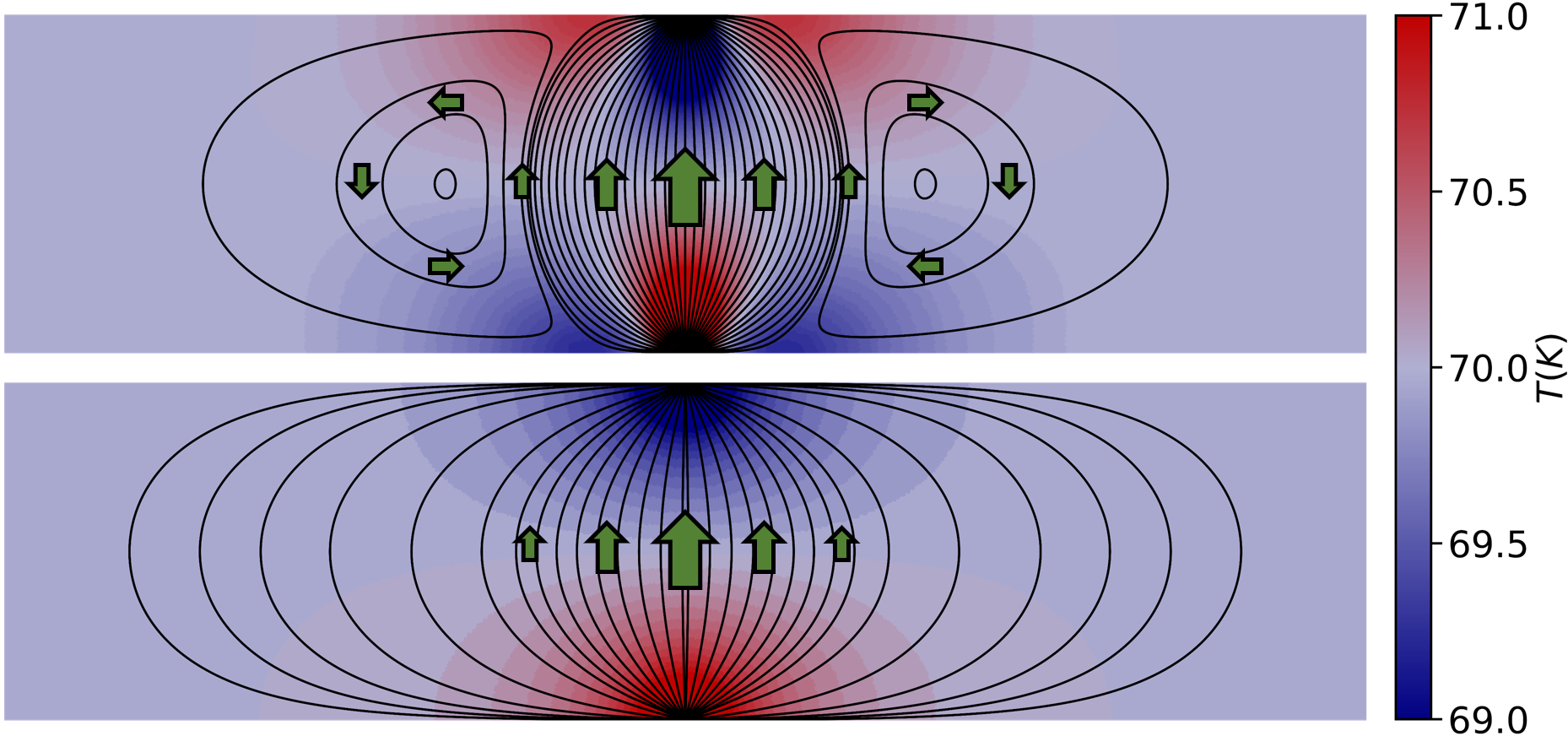}
\caption{Temperature map in the 2D graphite strip of width $y$ $=$ $h$, with streamlines shown for viscous (top) and diffusive (bottom) flow. In the viscous regime, negative thermal response arises from vortex-induced backflow opposing the main current, leading to negative thermal resistance across the strip. Streamlines (black lines) split into three regions: a central channel and two lateral flows. They correspond to isolines of Eq. \eqref{final_chi_xi_0}, while the temperature profile follows Eq. \eqref{T_xi_zero_and_epsilon_0}. These patterns serve as measurable signatures of compressible, vortical phonon flow. By contrast, diffusive heat transport aligns with the temperature gradient, yielding resistance along the flow as described by Eq. \eqref{final_temperature_Fourier}. Streamlines in this case extend outward from the central channel and directly connect source and drain, without forming vortices or nodal lines. Green arrows indicate flow direction along each streamline group. Boundary conditions are $\bar{T}$ $=$ $70$K, $\Delta T$ $=$ $1$K, and $U$ $=$ $500$m/s; remaining transport coefficients are chosen so that $\xi$, $\epsilon$ $\to$ $0$ for viscous flow and $\to$ $\infty$ for diffusive flow, emphasizing their qualitative differences.
}
\label{fig:fig1}
\end{figure}

Recent years have seen major theoretical \cite{lindsay2014phonon,cepellotti2015phonon,levitov2016electron,simoncelli2020generalization} and experimental \cite{bandurin2016negative,crossno2016observation,moll2016evidence,lee2015hydrodynamic,ding2022observation,huberman2019observation} advances in electrical and thermal transport in fast conductors, especially with the emergence of hydrodynamic regimes for electrons and phonons. Unlike conventional heat conduction, dominated by momentum-relaxing interactions, phonon hydrodynamics involves fluid-like heat flow driven by momentum-conserving phonon-phonon scattering. First studied in the 1960s, it revealed Poiseuille-like flow \cite{mezhov1965measurement} and second sound \cite{ackerman1966second,guyer1966solution,gurzhi1968hydrodynamic,enz1968one,hardy1970phonon,gotze1967first} in solid helium \cite{ackerman1966second}, sodium fluoride \cite{jackson1970second,pohl1976observation}, bismuth \cite{narayanamurti1972observation}, sapphire \cite{danil1979observation}, and strontium titanate \cite{hehlen1995observation}, all at cryogenic temperatures. Theoretical efforts bridged microscopic and macroscopic heat transport: Sussmann and Thellung \cite{sussmann1963thermal} derived mesoscopic equations from the linearized Boltzmann transport equation (LBTE) \cite{peierls1955quantum}, while Gurzhi \cite{gurzhi1964thermal,gurzhi1968hydrodynamic} and Guyer and Krumhansl \cite{guyer1966solution,guyer1966thermal} introduced weak momentum dissipation to model second sound and Poiseuille flow. Early models assumed specific phonon dispersions—linear-isotropic or power-law—but were later refined by Hardy's mesoscopic framework incorporating weak Umklapp scattering \cite{hardy1970phonon,hardy1974hydrodynamic}. Gurzhi’s work also paved the way for electron hydrodynamics, enabling recent observations of electron viscosity \cite{bandurin2016negative}. First-principles simulations with the LBTE predicted hydrodynamic behavior in graphene and other 2D materials \cite{cepellotti2015phonon,lee2015hydrodynamic,cepellotti2017transport}, carbon nanotubes \cite{lee2017hydrodynamic}, and graphite \cite{ding2018phonon} at non-cryogenic temperatures. Experiments have since confirmed viscous thermal transport in such conductors \cite{schmidt2008pulse,balandin2011thermal,fugallo2014thermal,machida2020phonon}, including room-temperature second sound \cite{melis2021room,beardo2021observation,ding2022observation,huberman2019observation}, Poiseuille-like flow \cite{huang2023observation,li2022reexamination,cepellotti2017boltzmann,machida2018observation,sendra2022hydrodynamic}, and lattice cooling \cite{jeong2021transient}. Interest in phonon hydrodynamics is growing, yet macroscopic viscous signatures remain elusive, and no definitive analytical detection method exists, limiting device-level applications.\\
The phonon LBTE is a key tool for simulating phonon hydrodynamics, relying on first-principles calculations of phonon lifetimes  \cite{cepellotti2016thermal,lee2015hydrodynamic,fugallo2013ab,chen2021non,di2023crossover}. Recently, thermal conductivity was reformulated as a sum over relaxons—collective phonon modes and scattering matrix eigenvectors \cite{cepellotti2016thermal}. Odd relaxons contribute to conductivity; even relaxons define thermal viscosity in the hydrodynamic regime. Coarse-graining the LBTE yields mesoscopic coupled partial differential equations—the viscous heat equations (VHE) \cite{simoncelli2020generalization}—analogous to the Navier-Stokes equations for laminar flow but with unique features. Besides temperature, they include a phonon fluid drift velocity, enabling a direct analogy with the pressure and velocity of classical fluids. The VHE describe hydrodynamic, diffusive, and intermediate regimes, greatly reduce computational cost versus full LBTE, and provide transparent physical interpretation. LBTE's limitations in handling complex geometries hinder practical predictions, whereas the VHE accommodate shape and boundary effects efficiently and have been benchmarked against spatially resolved LBTE results in micrometer-scale devices \cite{dragavsevic2023viscous}.\\
In this work, we show that the VHE can be decoupled into two modified biharmonic equations for the velocity potential and stream function, enabling an analytical solution of phonon hydrodynamics in Fourier space. We resolve the temperature profile and interpret it as a sum of vorticity and compressibility contributions. We also highlight the critical role of compressibility in viscous thermal transport—a defining feature of phonon fluids—unlike electronic fluids \cite{torre2015nonlocal,bandurin2016negative,levitov2016electron}, which are typically, though not always, incompressible \cite{levitov2016electron}. We identify the interplay between compressibility and vorticity as the origin of thermal viscosity and, mirroring Ref. \cite{levitov2016electron}, relate it to negative nonlocal thermal resistance in a 2D strip: a hallmark of viscous heat flow. This effect arises from thermal vortices \cite{raya2022hydrodynamic,restuccia2023non,sykora2023multiscale,shang2020heat,zhang2021heat,tur2024microscopic} or thermal backflow \cite{dragavsevic2023viscous}.\\
A compact form of the (steady-state and isotropic) VHE reads (see Supplementary Information (SI) \cite{supplementary} for details)
\begin{equation} \label{VHE}
\begin{cases}
\alpha\nabla\cdot\boldsymbol{u}=\kappa\nabla^{2}T\\
\beta\nabla T-\eta\nabla^{2}\boldsymbol{u}-\left(\zeta+\frac{\eta}{3}\right)\nabla\left[\nabla\cdot\boldsymbol{u}\right]=-\gamma\boldsymbol{u},
\end{cases}
\end{equation}
where $T$ and $\boldsymbol{u}$ denote temperature and phonon drift velocity fields, respectively; $\eta$ is the shear thermal viscosity, $\zeta$ the volume thermal viscosity, and $\kappa$ the thermal conductivity. Coupling coefficients $\alpha$ and $\beta$ derive from the energy–crystal momentum relation for phonons \cite{simoncelli2020generalization}, while $\gamma$ accounts for heat dissipation due to Umklapp and boundary scattering. The volume viscosity term vanishes for incompressible fluids, where flow divergence is zero. Thus, phonon fluids are generally compressible—unlike charge flow, which is mostly incompressible, provided that electronic drift velocities remain below plasmonic ones \cite{levitov2016electron}. Unless otherwise noted, VHE parameters are those computed from first principles in Ref. \cite{simoncelli2020generalization} for in-plane graphite with natural isotopic abundance.
\\
By defining a thermal compressibility $\Phi$ and vorticity $\boldsymbol{\mathcal{W}}$ respectively as
\begin{equation}
\begin{split}
&\Phi=\nabla\cdot\boldsymbol{u},\\
&\boldsymbol{\mathcal{W}}=\nabla\times\boldsymbol{u}
\end{split}
\end{equation}
we are able to decouple the VHE \eqref{VHE} and obtain a stand-alone modified Helmoltz equation for both $\Phi$ and $\boldsymbol{\mathcal{W}}$ \cite{supplementary}:
\begin{equation} \label{VHE_compressibility_vorticity}
\begin{cases}
\frac{\kappa\left(\eta+\mu\right)}{\beta}\nabla^{2}\Phi-\left(\frac{\kappa\gamma}{\beta}+\alpha\right)\Phi=0,\\
\eta\nabla^{2}\boldsymbol{\mathcal{W}}-\gamma\boldsymbol{\mathcal{W}}=0,
\end{cases}
\end{equation}
where $\mu=\zeta+\frac{\eta}{3}$. It is worth noting that the same equation for thermal vorticity holds also in the time-dependent regime \cite{supplementary}; this is not the case for thermal compressibility. The form \eqref{VHE_compressibility_vorticity} already allows for an analytical solution, but choosing appropriate boundary conditions for compressibility and vorticity is difficult due to the lack of direct experimental data on their behavior at the boundaries.
\\
To remedy this, we express the velocity vector via the velocity potential $\phi$ and the stream function $\Psi$ (see Eq. 18 in the SI \cite{supplementary}), using the Helmholtz decomposition \cite{batchelor1967introduction} to write it as a sum of curl-free and divergence-free components:
\begin{equation} \label{u_helmoltz_decomposition}
\boldsymbol{u}=-\nabla\phi+\nabla\times\boldsymbol{\Psi}=\left(-\frac{\partial\phi}{\partial x}+\frac{\partial\psi}{\partial y},-\frac{\partial\phi}{\partial y}-\frac{\partial\psi}{\partial x}\right),
\end{equation}
where, in the present case, $\psi$ is the only non-zero component of the stream function vector $\boldsymbol{\Psi}$, which points in the $z$-direction since we are focusing on a two-dimensional section. This enables solving the problem using the fields $\psi$ and $\phi$, with boundary conditions applied to $\boldsymbol{u}$. Using Eq. \eqref{u_helmoltz_decomposition}, we are able to recast the thermal compressibility and vorticity equations \eqref{VHE_compressibility_vorticity} into equations for $\phi$ and $\psi$ \cite{di2024vorticity,supplementary}:
\begin{equation} \label{potential_VHE_final}
\begin{cases}
\frac{\kappa(\eta+\mu)}{\beta}\nabla^{2}(\nabla^{2}\phi)-\left(\frac{\kappa\gamma}{\beta}+\alpha\right)\nabla^{2}\phi=0,\\
\eta\nabla^{2}(\nabla^{2}\psi)-\gamma\nabla^{2}\psi=0.
\end{cases}
\end{equation}
Note that the second equation in Eq. \eqref{potential_VHE_final} also holds for a general stream function $\boldsymbol{\Psi}$ in 3D, while the first remains unchanged since $\phi$ is always a scalar potential. Eqs. \eqref{VHE_compressibility_vorticity} and \eqref{potential_VHE_final} constitute the first main result of this work, showing that heat flow in a generic compressible phonon fluid—governed by temperature and velocity fields—can be fully decoupled into equations for vorticity and compressibility (or their associated potentials), allowing analytical solutions. Eqs. \eqref{potential_VHE_final} take the form of modified-biharmonic equations, generalizing the biharmonic equation for the stream function in incompressible Stokes flow \cite{batchelor1967introduction}, $\nabla^{2}(\nabla^{2}\psi)=0$, which is recovered in the limit $\gamma \to 0$.\\
As a case study, we examine thermal current in an infinitely long 2D strip device of width $h$ made of graphite \cite{levitov2016electron}.
We focus on the response to a temperature gradient and drift velocity injected and drained via point-like contacts, as shown in Fig. \ref{fig:fig1}. Two boundary conditions apply at the horizontal edges $y=0$ and $y=h$. The first enforces the drift velocity injected at the lower lead and extracted at the upper:
\begin{equation} \label{BCuy}
u_{y}(x,0)=u_{y}(x,h)=U\delta(x).
\end{equation}
The second imposes a no-slip condition for the tangential velocity:
\begin{equation} \label{BCux}
u_{x}(x,0)=u_{x}(x,h)=0.
\end{equation}
Temperature boundary conditions assume $T_{{\rm{hot}}}$ and $T_{{\rm{cold}}}$ baths centered at $y=h$ and $y=0$. To match the injected velocity from $y=0$ to $y=h$, we impose:
\begin{equation} \label{temperature_BC_y=0,h}
T(x,0)=\bar{T}+\Delta T\delta(x),\,\,\,\,\,T(x,h)=\bar{T}-\Delta T\delta(x)
\end{equation}
producing a gradient centered at the equilibrium temperature $\bar{T}$ along the drift direction (the experimental setup for these boundary conditions is discussed in the SI \cite{supplementary}). In the strip geometry depicted in Fig. \ref{fig:fig1}, the problem can be conveniently analyzed in a mixed position-momentum representation via Fourier transforms.
By exploiting the symmetry of the phonon drift velocity components ($u_{x}(x,y)=-u_{x}(x,h-y)$ and $u_{y}(x,y)=u_{y}(x,h-y)$) the solutions of Eq. \eqref{potential_VHE_final} read \cite{supplementary}
\begin{equation} \label{solution_psi_phi}
\resizebox{0.5\textwidth}{!}{$
\begin{cases}
\psi(k,y)=a_{\psi}(k)\left(e^{ky}+e^{kh}e^{-ky}\right)+b_{\psi}(k)\left(e^{q_{\psi}y}+e^{q_{\psi}h}e^{-q_{\psi}y}\right)\\
\phi(k,y)=a_{\phi}(k)\left(e^{ky}-e^{kh}e^{-ky}\right)+b_{\phi}(k)\left(e^{q_{\phi}y}-e^{q_{\phi}h}e^{-q_{\phi}y}\right),
\end{cases}$}
\end{equation}
where $q_{\psi}^{2}=k^{2}+\frac{\gamma}{\eta}$ and $q_{\phi}^{2}=k^{2}+\frac{\gamma}{\eta+\mu}+\frac{\alpha\beta}{\kappa(\eta+\mu)}$.
\\
We now examine the temperature profile of viscous flow, obtained by evaluating Eq. \eqref{VHE} using Eq. \eqref{u_helmoltz_decomposition} and substituting Eq. \eqref{solution_psi_phi}, yielding
\begin{equation} \label{T_as_sum_of_T_phi_and_T_psi}
\begin{split}
T(x,y)=T_{\phi}(x,y)+T_{\psi}(x,y),
\end{split}
\end{equation}
where
\begin{equation} \label{T_profile}
\begin{split}
&T_{\phi}(x,y)=\frac{1}{2\pi}\int dke^{ikx}\Bigg[\frac{\gamma}{\beta}a_{\phi}(k)\left(e^{ky}-e^{kh}e^{-ky}\right)-\\
&\hspace{3.75cm}-\frac{\alpha}{\kappa}b_{\phi}(k)\left(e^{q_{\phi}y}-e^{q_{\phi}h}e^{-q_{\phi}y}\right)\Bigg],\\
&T_{\psi}(x,y)=\frac{i}{2\pi}\int dke^{ikx}\frac{\gamma}{\beta}a_{\psi}(k)\left(e^{ky}-e^{kh}e^{-ky}\right).
\end{split}
\end{equation}
This represents the second main result of this work: the temperature profile of a phonon fluid can be decomposed into two components, associated with the curl and divergence of the drift velocity. Crucially, this decomposition is general and not limited to the strip geometry. To analyze the diffusive and viscous limits of Eq. \eqref{T_profile}, we analytically recover the dimensionless Fourier deviation number (FDN) \cite{simoncelli2020generalization}, which quantifies deviations from Fourier’s law due to hydrodynamic effects:
\begin{equation} \label{FDN}
\text{FDN}=\frac{1}{\epsilon+\xi},\text{ where }\epsilon=\frac{\gamma h^{2}}{\eta}\text{ and }\xi=\frac{\kappa}{\alpha}\frac{\frac{\Delta T}{h}}{U}.
\end{equation}
Larger FDN implies stronger deviations from diffusive behavior. The $\epsilon$ term increases FDN as $\gamma$ decreases, i.e., when crystal momentum dissipation via Umklapp scattering weakens. Similarly, FDN grows as $\xi$ decreases, meaning the injected phonon drift velocity $U$ dominates over conductivity and temperature gradient. When momentum-dissipating scattering dominates, the temperature profile reduces to the diffusive limit described by Fourier’s law:
\begin{equation} \label{final_temperature_Fourier}
T(x,y)=\frac{1}{2\pi}\int dk\,e^{ikx}\frac{T_{{\rm{bc}}}}{1-e^{kh}}\left(e^{ky}-e^{kh}e^{-ky}\right),
\end{equation}
with reciprocal-space boundary condition $T_{{\rm{bc}}}=\bar{T}\delta(k)+\Delta T$. This expression is used to plot the temperature map in Fig. \ref{fig:fig1}b.\\
We now analyze the strip’s temperature profile in the ideal hydrodynamic regime, starting with the thermal response in the small $\xi$ limit.
In this case we have \cite{supplementary}
\begin{equation} \label{T_xi_to_zero}
\resizebox{0.475\textwidth}{!}{$
\begin{split}
&T_{\phi}(x,y)=\frac{1}{2\pi}\int dk\,e^{ikx}\Bigg[\frac{T_{{\rm{bc}}}}{1-e^{kh}}\left(e^{ky}-e^{kh}e^{-ky}\right)+\frac{1-e^{kh}}{1-e^{q_{\phi}h}}\cdot\\
&\hspace{3.75cm}\cdot\frac{U\gamma}{\beta}\frac{q_{\psi}}{k}G(q_{\psi},k)\left(e^{q_{\phi}y}-e^{q_{\phi}h}e^{-q_{\phi}y}\right)\Bigg],\\
&T_{\psi}(x,y)=\frac{1}{2\pi}\frac{\gamma}{\beta}U\int dk\,e^{ikx}\frac{q_{\psi}}{k}G(q_{\psi},k)\left(e^{ky}-e^{kh}e^{-ky}\right),
\end{split}$}
\end{equation}
where 
\begin{equation} \label{condensed_notation_xi}
G(q_{\psi},k)=\frac{1-e^{q_{\psi}h}}{q_{\psi}(1-e^{q_{\psi}h})(1+e^{kh})-k(1+e^{q_{\psi}h})(1-e^{kh})}.
\end{equation}
Eq. \eqref{T_xi_to_zero} predicts an inverse-square dependence versus distance from the injection points and also the presence of two nodal lines along the directions $y=x$ and $y=-x$ related to the main vertical heat path. This can be seen looking at the viscous regime where $(q_{\psi}-k)\to0$ (that means $\epsilon\to0$), leading to
\begin{equation} \label{T_xi_zero_and_epsilon_0}
\resizebox{0.475\textwidth}{!}{$
\begin{split}
T(x,y)&=\frac{1}{2\pi}\int dk\,e^{ikx}\Bigg[\sinh\left(ky-k\frac{h}{2}\right)\cdot\\
&\hspace{2.6cm}\cdot\left(-\frac{8U\frac{\eta}{\beta}k\sinh\left(k\frac{h}{2}\right)}{kh+\sinh(kh)}-\frac{T_{{\rm{bc}}}}{\sinh\left(k\frac{h}{2}\right)}\right)\Bigg]\approx\\
&\approx\frac{4U}{\pi}\frac{\eta}{\beta}\frac{x^{2}-y^{2}}{(x^{2}+y^{2})^{2}}-\frac{\Delta T}{\pi}\frac{y}{x^{2}+y^{2}}.
\end{split}$}
\end{equation}
As seen in Fig. \ref{fig:fig1}a, the resulting spatial dependence of this temperature profile features a remarkable behavior.
\begin{figure}[!htb]
\centering
\includegraphics[width=0.48\textwidth]{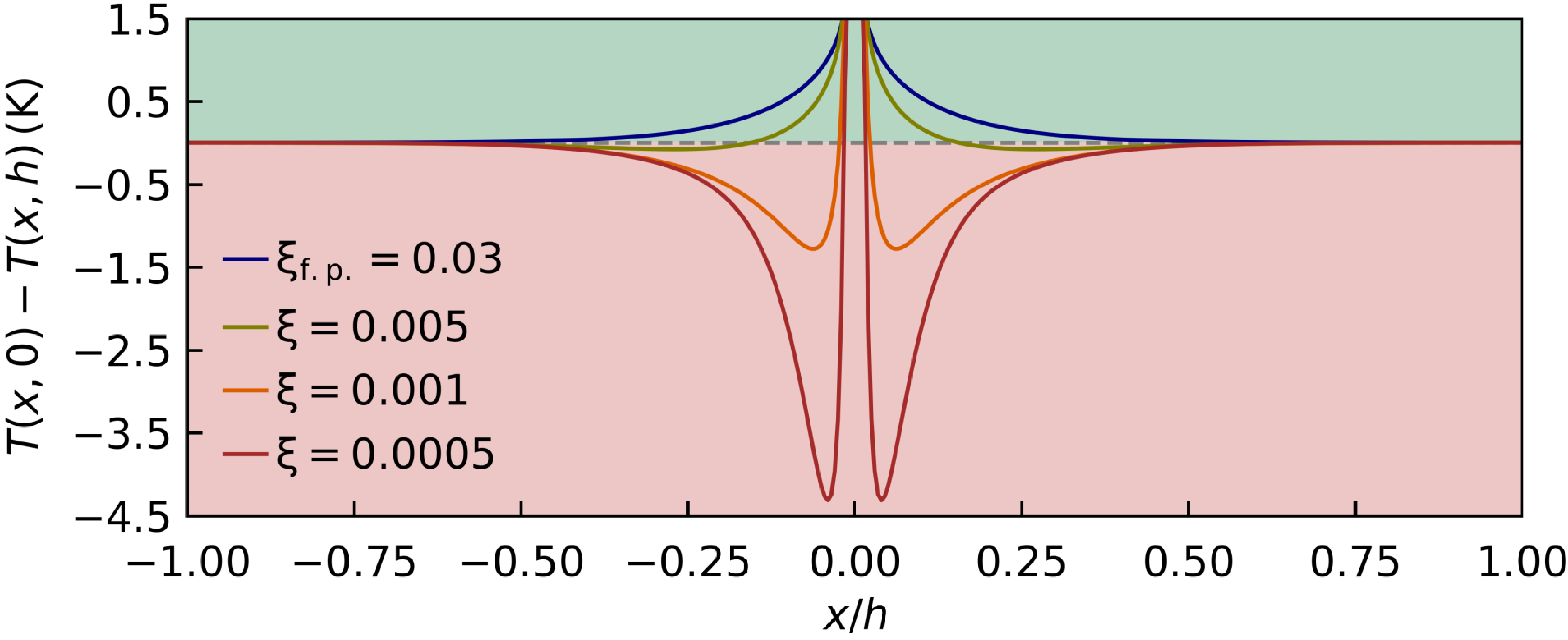}
\caption{Non-local thermal response in viscous and diffusive regimes. The temperature difference $T(x,0)$ $-$ $T(x,h)$ is plotted versus distance $x$ from the leads for various $\xi$ at fixed $\epsilon$ $=$ $0.1$ (see Eq. \eqref{temp_diff_xi_bis}). In the diffusive region (large $|x|$), the response is positive, while near the leads—where viscosity dominates—it becomes negative. Positive values at very small $|x|$ reflect the finite contact size ($\sim$$0.05h$) used in simulations (see main text). Fourier-like behavior persists up to relatively high $\xi$; $\xi_{\rm{f.p.}}$ denotes the first-principles value for in-plane graphite with natural isotopic abundance. Green and pink areas indicate positive and negative response, respectively.}
\label{fig:fig2}
\end{figure}
Temperature reaches equilibrium along $y=h/2$ by symmetry, showing multiple sign changes with nodal lines separating $T>\bar{T}$ and $T<\bar{T}$ regions. For small $x$ near the thermal path $x=0$, temperature shifts from positive at the source to negative at the drain, relative to center equilibrium. This pattern reverses away from the main path. At $y=0,h$, outside the leads, the temperature difference sign opposes that at the leads. This follows from what
happens at the borders between the nominal thermal flow path region and adjacent regions where thermal vortices arise (vortex centers near $x=\pm h$, per numerical analysis). Crystal momentum diffuses transversely, diverging from the central path. Consequently, any point outside the leads near the upper boundary connects to a symmetrical point near the lower via a streamline opposing central flow (see Eq. \eqref{final_chi_xi_0}), implying negative thermal resistance. Moreover, the phonon drift velocity part in Eq. \eqref{T_xi_zero_and_epsilon_0} mimics the solution in Ref. \cite{levitov2016electron} for the electrical potential. Near the leads ($x$ $\to$ $0$), the drift velocity term decays as $-y^{-2}$, dominating over the temperature gradient term ($\sim$$-y^{-1}$) when approaching the boundaries.\\
The temperature difference across the strip obtained using Eqs. \eqref{T_xi_to_zero} is
\begin{equation} \label{temp_diff_xi}
\resizebox{0.475\textwidth}{!}{$
T(x,0)-T(x,h)=\frac{1}{\pi}\int dk\,e^{ikx}\left[\Delta T+\frac{2U\gamma}{\beta}\left(1-e^{kh}\right)\frac{q_{\psi}}{k}G(q_{\psi},k)\right]$}.
\end{equation}
In the limit of $(q_{\psi}-k)\to0$, Eq. \eqref{temp_diff_xi} reads \cite{supplementary}
\begin{equation} \label{temp_diff_xi_bis}
\resizebox{0.475\textwidth}{!}{$
\begin{split}
T(x,0)-T(x,h)&\approx\frac{1}{\pi}\int dk\,e^{ikx}\left[\Delta T+4U\frac{\eta}{\beta}\frac{k\tanh\left(k\frac{h}{2}\right)\sinh(kh)}{kh+\sinh(kh)}\right]\to\\
&\xrightarrow[|k|\gg\frac{1}{h}]{}2\Delta T\delta(x)-8U\frac{\eta}{\beta}\frac{1}{x^{2}}.
\end{split}$}
\end{equation}
This is consistent with the result for electron fluids in Ref. \cite{levitov2016electron}. Fig. \ref{fig:fig2} shows the thermal response across the strip for various $\xi$. At high $\xi$, the response is positive (diffusive regime), while at low $\xi$—near the leads—the thermal resistance becomes negative, as prescribed by Eq. \eqref{temp_diff_xi_bis}. This negative response in the low $\xi$ limit emerges from reduced conductivity and temperature gradient, and especially from increased drift velocity. In contrast, tuning momentum dissipation via $\epsilon$ (Umklapp scattering) has a much smaller effect on backflow (not shown). Thus, a sufficiently low $\xi$ is essential for thermal backflow, whereas reducing momentum dissipation alone is insufficient. This can be more easily understood when discussing the incompressible and irrotational limits.\\
The thermal response behavior is strongly affected by the injected drift velocity $U$, which in fact takes part into the definition of $\xi$. To isolate its influence, we fix all transport coefficients to their first-principles values for in-plane graphite with natural isotopic abundance \cite{simoncelli2020generalization}, and vary only $U$ (Fig. \ref{fig:fig3}). For a 1K temperature gradient, a drift velocity $\gtrsim20000$m/s is required to produce even a 0.1K thermal backflow, posing a challenge for experiments. Using isotopically purified graphite ($^{12}$C 99.95\%, $^{13}$C 0.05\%) boosts the backflow to 0.4K.
\\
The robustness of the negative thermal response can be explained by the fact that viscosity is associated with the Laplacian (second derivative) of the velocity in the second term of Eq. \eqref{VHE}, making it dominant at short distances. As Fig. \ref{fig:fig2} shows, this may be key to experimentally observing viscous thermal transport. Note that, following Ref. \cite{levitov2016electron}, we replace the Dirac deltas in boundary conditions \eqref{BCuy} and \eqref{temperature_BC_y=0,h} with Lorentzians, causing sign changes at the contact edges (equal to the HWHM) and negative response outside the contacts.\\
To better understand the origin of the vortices in Fig \ref{fig:fig1}a, we compute the flow streamlines. In the present case of a general compressible and rotational phonon fluid, these are given by the condition $\chi(x,y)=\rm{const.}$ \cite{batchelor1967introduction}, where
\begin{equation} \label{complex_potential}
\chi(x,y)=i\psi(x,y)-\phi(x,y)
\end{equation}
is the so-called complex potential \cite{batchelor1967introduction,orlofftopic} of the flow. Here, the flow is described by a drift velocity vector (see Eq. \eqref{u_helmoltz_decomposition}) with components that cannot be represented solely as derivatives of a single scalar field $\psi$, as in incompressible flows. Instead, they also involve derivatives of $\phi$, which accounts for compressibility. So in this case both compressibility and vorticity define the nature of the streamlines in Fig. \ref{fig:fig1}a. These are derived from the $\xi\to0$ solution of the complex potential. By taking the limits as $(q_{\psi}-k)\to0$ and $(q_{\phi}-k)\to0$, we obtain
\begin{equation} \label{final_chi_xi_0}
\resizebox{0.475\textwidth}{!}{$
\begin{split}
\chi_{\xi\to0}(x,y)\sim\frac{U}{2\pi}\int dk\,e^{ikx}\frac{1}{k}\Bigg[&\frac{\cosh\left[k\left(y-\frac{h}{2}\right)\right]}{\cosh\left(k\frac{h}{2}\right)}+\frac{k\tanh\left(k\frac{h}{2}\right)}{kh+\sinh(kh)}\cdot\\
&\cdot\Big[y\sinh[k(h-y)]+(h-y)\sinh(ky)\Big]\Bigg].
\end{split}$}
\end{equation}
Although the majority of streamlines are open lines connecting source ($y=0$) and drain ($y=h$), some of them form loops (see Fig. \ref{fig:fig1}a), leading to the vortices occurring on both sides of the current path.
\begin{figure}[h]
\centering
\includegraphics[width=0.475\textwidth]{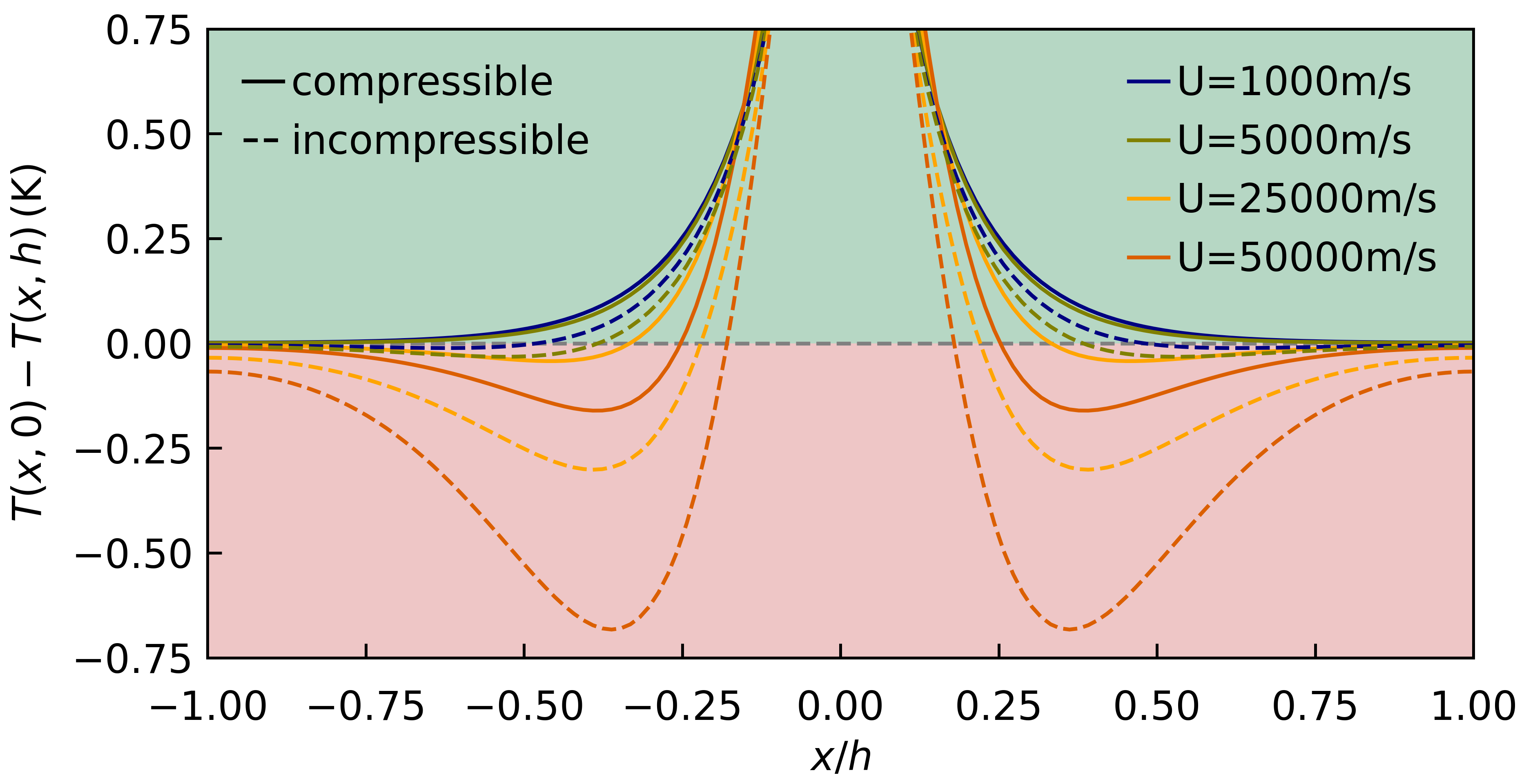}
\caption{Comparison of the non-local thermal response between the compressible and incompressible cases. The temperature difference $T(x,0)-T(x,h)$ is graphed as a function of distance $x$ from the leads for different values of the injected phonon drift velocity $U$. Full lines represent the general compressible case while dashed lines represent the incompressible limit (see SI \cite{supplementary}). Blue and pink regions mark positive and negative thermal response, respectively. Here $\Delta T=1$K and the transport coefficients of in-plane graphite are those computed from first principles in Table I of Ref. \cite{simoncelli2020generalization}.}
\label{fig:fig3}
\end{figure}
These streamlines mirror those of the electron fluid obtained in Eq. 8 of the SI of Ref. \cite{levitov2016electron}.\\
At this stage, we can isolate the irrotational and incompressible components of the temperature in Eq. \eqref{T_as_sum_of_T_phi_and_T_psi}, clarifying which contributions primarily drive the vortices and the resulting thermal backflow. In the irrotational limit ($\boldsymbol{\mathcal{W}}=\nabla\times\boldsymbol{u}=0$), the drift velocity derives from the single scalar potential $\phi$, $\boldsymbol{u}=-\nabla\phi=(-\frac{\partial\phi}{\partial x}, -\frac{\partial\phi}{\partial y})$, making the second of Eqs. \eqref{potential_VHE_final} irrelevant. Solving the first of Eqs. \eqref{potential_VHE_final} with boundary conditions \eqref{BCuy} and \eqref{BCux}, one finds that injecting drift velocity alone does not induce thermal flow \cite{supplementary}. However, allowing for a temperature gradient, the irrotational temperature profile reads
\begin{equation} \label{irr_temp}
T^{\boldsymbol{\mathcal{W}}=0}\equiv T_{\phi}^{\epsilon\to0}.
\end{equation}
On the other hand, in the incompressible limit ($\Phi=\nabla\cdot\boldsymbol{u}=0$), the velocity derives from a single stream function $\psi$, $\boldsymbol{u}=\nabla\times\boldsymbol{\Psi}=(\frac{\partial\psi}{\partial y}, -\frac{\partial\psi}{\partial x})$, making the first of Eqs. \eqref{potential_VHE_final} irrelevant. When solving the second of Eqs. \eqref{potential_VHE_final} considering boundary conditions \eqref{BCuy} and \eqref{BCux}, the incompressible temperature profile becomes
\begin{equation} \label{inc_temp}
T^{\Phi=0}\equiv T_{\psi}^{\xi\to0}.
\end{equation}
This is shown \cite{supplementary} to have the same form as the electrical potential in Ref. \cite{levitov2016electron}, where charge transport is described by an incompressible flow. Thus, like electron fluids, it is possible to generate thermal flow in the device by injecting drift velocity, as the phonon flow is incompressible. Eqs. \eqref{irr_temp} and \eqref{inc_temp} show that in the limit $\epsilon \to 0$, the temperature profile corresponds to the irrotational part of the solution, while $\xi \to 0$ represents the incompressible component. In Fig. \ref{fig:fig3}, we evaluate the incompressible thermal response across the strip and compare it to the general compressible case. The results demonstrate that thermal backflow through viscous vortices is favored in incompressible phonon fluids. This analysis highlights how $T_{\psi}$ plays a key role in thermal vortices and backflow, as it is linked to $\xi$, which dominates the viscous behavior. So, we see that the temperature profile in Eq. \eqref{T_as_sum_of_T_phi_and_T_psi} can be interpreted as the sum of two components, corresponding to the irrotational and incompressible contributions in the viscous limit ($\rm{FDN \to \infty}$).\\
In summary, the present study shows how to separate the steady-state VHE into modified biharmonic equations for the velocity potential and stream function of the phonon fluid. We derive analytical solutions in Fourier space, define streamlines using a complex potential, and reveal temperature as a sum of compressibility and vorticity contributions.
Notably, we show how heat current can backflow against the injected one, highlighting the non-local thermal response in viscous flow. Exploring irrotational and incompressible limits, we identify incompressibility as essential for strong backflow, offering a practical criterion for selecting systems with pronounced viscous behavior. The present approach also applies to electron fluids where drift exceeds plasmon velocity, violating incompressibility. Finally, we propose an experimental setup (see SI \cite{supplementary}) for detecting thermal backflow in in-plane graphite and other 2D systems. Materials such as graphene, graphite, and diamond, easily patterned without degrading their properties, are promising candidates for phononic and electronic microfluidics.\\
E.D. acknowledges support from the Swiss National Science Foundation (SNSF), through Grant No. CRSII5\_189924 (“Hydronics” project). E.D. also thanks Michele Simoncelli for fruitful discussions and insights during the course of this work. N.M. acknowledges support from NCCR MARVEL, a National Centre of Competence in Research, funded by the Swiss National Science Foundation (Grant No. 205602).\\

\bibliographystyle{apsrev4-1}

%

\end{document}